\magnification=\magstep1
\hsize 32 pc
\vsize 42 pc
\baselineskip = 24 true pt
\def\cl{\centerline}
\def\vs {\vskip .5 true cm}
\cl {\bf Fabrication of Magnetic Charge from Excited States of H-atom}
\vs
\cl {\bf T. PRADHAN}
\cl {Institute of Physics, Bhubaneswar-751005, INDIA}
\vs
\cl { PACS - 14.80 Hv}
\vs
\cl { Abstract}

It is shown that the excited states of hydrogen atom in a uniform
electric field (Stark States) posess magnetic charge whose magnitude is
given by a Dirac-Saha type relation:
$$ {eg\over \hbar c} = \sqrt 3 n $$
An experiment is proposed to fabricate such states and to detect 
their magnetic charge.
\vfill
\eject
Dirac [1] incorporated magnetic charge (monopole) in electrodynamics by introducing
a string singularity in the vector potential and showed that its strength is given 
by the relation
$$ {eg\over \hbar c} = {n\over 2}, \ n = 1,2,3...\eqno{(1)}$$ 
This relation was independently obtained by Saha [2] and Wilson [3]   by quantizing the angular 
momentum of a two-body system consisting of a magnetic point charge and an
electric point charge. Existence of magnetic charge is also a feature of certain  non-abelian gauge theories [4]. 
Although numerous experimental searches [5] have been undertaken to detect them, 
none have been found so far. In this note we show that it is possible to fabricate
magnetically charged states from the excited states of the hydrogen atom by
putting it in a uiform external electrostatic field and detect their magnetic
charge by a simple experiment.

If magnetic charges exist, Maxwell equations take the form:
$$\eqalign {\vec\nabla .\vec E &  = \rho^{(e)}, \ \
\vec\nabla\times\vec B - {\partial\vec E\over 
\partial t} =   \vec J^{(e)}\cr 
\vec\nabla .\vec B &  = \rho^{(m)}, \ \ \vec\nabla\times\vec E + {\partial\vec B\over
\partial t} = - \vec J^{(m)}\cr} \eqno{(2)}$$
where the superscripts (e) and (m) on $\vec J$ and $\rho$ stand for
electric and magnetic densities. Our contention is that 
analogous to the relation
$$ \vec\mu = {1\over 2} \int d^3 r (\vec r\times \vec J^{(e)}  (\vec
r)  \eqno{(3)}$$
between magnetic dipole moment $\vec\mu$ and electric current $\vec
J^{(e)}$, there      a the relation
$$\vec d = {1\over 2} \int d^3 r [\vec r\times \vec J^{(m)} {(\vec r)}] 
\eqno{(4)}$$
between the permanent electric dipole moment $\vec d$ and the
magnetic current $ \vec J^{(m)}$. In quantum mechanics this relation
takes the form 
$$ <\vec d> = {1\over 2} \int d^3 r \  \psi^* {(\vec r)} [\vec r\times
\vec J^{(m)}  (\vec r)  ]\psi(\vec r) \eqno{(5)}$$
 Since $[\vec r\times \vec J^{(m)}  (\vec r) ]$ has odd parity, $<\vec 
d>$ will vanish since in general $\psi(\vec r)$ has definite 
parity. However, excited states of hydrogen atom do not have definite 
parity on account of l-degeneracy. These states have permanent
electric dipole moment.

It may be noted here that the electric dipole moment is usually 
defined as 
$$\vec d =\int d^3 r \  \vec r \rho^{e)}(\vec r)\eqno{(6)}$$
>From Maxwell equations (2), this can be written as
$$d_i = \int d^3 r \  r_i \partial_j E_j = - \int d^3 r \  E_i+\int d^3 r \
\partial_j (r_iE_j)\eqno{(7)}$$ 
The second term on the right can be converted into a surface integral 
which for the surface at infinity would vanish on account of the fact 
that the expectation value of electric field $\vec E$ produced by the 
charge distribution of the hydrogen atom vanishes exponentially at  
infinity. Similarly the definition (4) for the electric dipole moment 
can be written as  
$$ d_i=  {1\over 2} \int d^3 r(\vec r\times \vec J^{(m)}) = {1\over 2}
\int d^3 r \  r_j  (-\partial_iE_j+\partial_jE_i) = $$
$$ - \int d^3 r E_i +{1\over
2}  \int d^3 r                    
[\partial_j(r_jE_i)-\partial_i(r_jE_j)]\eqno{(8)}$$
in the static limit i.e. ${\partial\vec B\over \partial t} =0$. Since
the surface integrals vanish for hydrogen atom, we have the identity 

$$ \int d  ^3 r \ \vec r \rho^{(e)} (\vec r) = {1\over 2 } \int d^3 r   [\vec r\times
\vec J^{(m)} (\vec r)]\eqno{(9)}$$
Physically what this means is that the motion of the electron in the
hydrogen atom which gives rise to a permanent electric dipole moment
can be considered to have generated a magnetic current in terms of
which this dipole moment can be expressed.

>From parity and dimensional considerations the magnetic current $\vec
J^{(m)}(\vec r)$ can be written as
$$\vec J^{(m)}(\vec r) = g {\vec L\over mcr}\eqno{(10)}$$
where $\vec L$ is the orbital angular momentum and g is the magnetic
charge. The quantum machenical electric dipole moment operator in that case
can be written as
$$\vec d = {1\over 2} {\vec r\times\vec L\over mcr} = -{i\hbar g\over
2\times 2mcr} (\vec r\times \overline\nabla)\eqno{(11)}$$
$$where  \ \    a \overline\nabla b = a(\vec \nabla b)- (\vec \nabla a)b$$
 When the
hydrogen atom is placed in a static  uniform electric field $ (0,0,{\cal E}$) the
interaction energy operator $H_z$ has the form 
$$ H_I =\vec d. \vec {\cal E} = {-ig\lambda{\cal E}   \over 2\times 2} [x^2+y^2) 
{\overline\partial\over \partial z}
-z (x{\overline \partial\over\partial x}-y{\overline\partial
\over \partial y})]\eqno{(12)}$$
where $\lambda ={\hbar\over mc}$ is the compton wavelength of the
electron.
In parabolic co-ordinates
$$H_I = {-ig\lambda {\cal E}\over 2} {\xi\eta\over\xi+\eta}
({\overline\partial\over
\partial\xi}-{\overline\partial\over\partial\eta})\eqno{(13)}$$
The Schroedinger equation takes the form 
$$(H_0+H_I)\Phi = E\Phi = (E_0+\Delta  E)\Phi\eqno{(14)}$$
with
$$H_0 = {-\hbar^2\over 2m} \{ {4\over \xi+\eta} [{\partial\over\partial\xi}
(\xi{\partial\over\xi})+{\partial\over\partial \eta} (\eta
{\partial\over
\partial\eta})] +{1\over \xi\eta}
{\partial^2\over\partial\varphi^2}\}-{2e^2\over\xi+\eta}\eqno{(15)}$$
In order to solve eqn(14) we take
$$        E= E_0 +\Delta E, \ \Phi = \psi exp                          
({ieg\over \hbar c} F-iM\varphi) \ \ \ 
 F = log {\xi\eta\over a^2_0}  \eqno{(16)}$$
where $a_0$ ground state Bohr radius, M is magnetic quantum number
and $\psi$ is real. To order g
$$H_0\Phi =        E_0\Phi\eqno{(17)}$$
becomes
\vfill
\eject
$$-{\hbar^2\over 2m} ({4\over \xi+\eta}) [{\partial\over\partial\xi}
(\xi{\partial\psi\over \partial \xi} +{\partial\over
\partial\eta}(\eta{\partial\psi\over\partial \eta})] +{ieg\over \hbar c}
[{\partial\over \partial\xi} (\xi{\partial F\over\partial \xi})
+{\partial\over\partial\eta} (\eta{\partial F\over\partial \eta})] $$
$$ + {2ieg\over\hbar c} [{\partial\psi \over \partial\xi} (\xi{\partial
F\over\partial \xi}) +{\partial\psi \over\partial\eta} (\eta{\partial
F\over\partial \eta})]+{\hbar^2M^2\over 2m} {\psi \over\xi\eta} -
{-2e^2\psi\over
\xi+\eta} \psi = E_0\psi\eqno{(18)}$$
With
$$ F = log {\xi\eta\over a_0^2} \ , \ \ {\partial F\over\partial\xi} =
{1\over\xi} , \ \ \  {\partial\over\partial\xi} (\xi{\partial F\over
\partial\xi})=0\eqno{(19)}$$
equation (18) becomes
$$-{\hbar^2\over 2m} ({4\over \xi+\eta}) [{\partial\over\partial\xi}
(\xi{\partial\psi\over \partial \xi} +{\partial\over
\partial\eta}(\eta{\partial\psi\over\partial \eta})] +{2ieg\over \hbar c}
({\partial\psi\over\partial\xi}+{\partial\psi\over\partial\eta})]
+{\hbar^2M^2\psi\over 2m\xi\eta} -{2c^2\psi\over \xi+\eta} =
E_0\psi \eqno{(20)}$$
Since $E_0$ is obtained by multiplying both sides of this equation on
the left by $\psi^*$ and taking volume integral and since $\psi^*=\psi$
$$\int^{\infty}_0 d\xi\psi^* {\partial\psi\over\partial\xi} = {1\over 2}
\int^{\infty}_0 d\xi {\partial(\psi\psi)\over \partial\xi} =
0 $$
equation (20) is equivalent to
$$-{\hbar^2\over 2m} ({4\over \xi+\eta}) [{\partial\over\partial\xi}
(\xi{\partial\psi\over \partial \xi} +{\partial\over
\partial\eta}(\eta{\partial\psi\over\partial \eta})] +{\hbar^2M^2  \over
2m} {\psi\over \xi\eta} -{2e^2\psi\over\xi+\eta} = E_0\psi
\eqno{(21)}$$
which is the standard H-atom Schroedinger equation in the absence of external
field ${\cal E}$.

>From eqn(13) (14) and (16) we have, to orger $g^2$,
$$\Delta E = <\Phi\mid H_I\mid\Phi> = {ig\lambda {\cal E}  \over 8} \int
d\xi d\eta  \xi\eta \psi
({\overline\partial\over\partial\xi}-{\overline\partial\over
\partial\eta}) \psi - {eg^2\lambda {\cal E} \over 4\hbar c} \int d\xi
d\eta \psi ({\partial F\over\partial \xi}-{\partial F\over \partial
\eta})\psi \eqno{(22)}$$
which on use of (19) gives

$$\Delta E = {eg^2\lambda {\cal E}\over 4\hbar c}\int  d\xi d\eta\psi
(\xi-\eta) \psi= {eg^2\lambda\over 2\hbar c}\epsilon ({n_1-n_2\over
n})\eqno{(23)}$$
on account of vanishing of the first term on r.h.s. of eqn(22). The
conventional approach with the definition (6) for electric dipole
moment, gives
$$\Delta E = {3 \eta {\cal E} \over 2} (n_1-n_2) {\hbar^2\over
me}\eqno{(24)}$$
In view of the identify (9) we equate (23) and (24) and obtain
$${eg\over \hbar c} = \sqrt 3 n\eqno{(25)}$$
which differs from Dirac's    result by a numerical factor. 

It is worth noting that the wave function $\Phi$ in eqn(16) has a
logarithmic singularity either for $\xi= r+z=0$ or for $\eta = r-z =
0$. The former corresponds to singularity extending from 0 to
$-\infty$ along the negative z-axis and the latter extending from 0 to 
$+\infty$  along the positive z-axis which is reminiscent of the Dirac
string.

The linear combination of degenerate excited states of hydrogen atom
that posses permanent electric dipole moment and have magnetic charge
are just those discussed in this paper i.e. states with the parabolic
co-ordinate. These states form when the hydrogen atom is placed in an
uniform static electric field. Magnetic charged states can therefore
be experimentally  fabricated by passing hydrogen atoms selectively
excited to a specified n quantum number through a region of uniform
electrostatic field. For convenience one can excite the atoms to the n 
= 2 state  by absorption of a laser beam of $\lambda=2430  A$ which
can be generated by frequency doubling of an intense dye laser beam at 
$\lambda = 4860 \ A$ [6]. The two states of the beam with quantumnumbers $(n_1=0, \  n_2 = 1, \  m = 0)$ and $(n_1=1, n_2=0, m=0)$ possesing
equal and opposite magnetic charge get bent in opposite directions due 
to the action of the electrostatic field analogous to the bending of
electrically charged particles in uniform magnetostatic field. The two separated magnetic charged particle beams can be
passed through superconducting rings [4] causing flow of electric
current therein which can be detected by SQUID (Superconducting
quantum interpreference device).

In view of the fact that the non-degenerate ground state $\psi_o$ of
the hydrogen atom can have  no magnetic charge, it would appear that
conservation of magnetic charge will be violated when the atom makes a 
radiative transition to (form) the ground state from (to) an excited
state. However, a closer analysis shows that this is not so. Due to
interaction of the atom with the radiation field, the states acquire
time dependence as per below.
$$\eqalign{ \psi_n(t) & = [ Cos \  \omega_n t] \ \psi_n (0) + \ [ Sin 
\ \omega_n t]
\psi_0 (0)\cr
 \psi_0(t) & = [ Cos \  \omega_n t] \ \psi_0 (0) -  \ [ Sin \  \omega_n t]
\psi_n (0)\cr}\eqno{(26)}$$
where $\omega_n$ is the frequency of radiation resulting from the
transition between states $\mid n >$ and $\mid 0 >$. It    therefore
follows that
$$ \eqalign { g_n (t) & = g_n \ Cos^2 \omega_n \ t\cr
 g_0 (t) & = g_n \ Sin^2 \omega_n \ t\cr} \eqno{(27)}$$
so that
$$ g_n (t) + g_0 (t) = g_n\eqno{(28)}$$
 remains constant.

\vfill
\eject
\centerline {\bf References}
\vs 
\item {1.} P.A. M. Dirac, Proc. Roy. Soc. {\bf A133} 60 (1931), Phys.
Rev. {\bf 74} 817 (1948).
\item {2.} M.N. Saha, Ind. J. Phys. {\bf 10} 141 (1936), Phys. Rev. {\bf 75} 1968 (1949).
\item {3.} G't Hooft, Nucl. Phys. {\bf B79} 276 (1974).
\item {} A.M. Polyakov, JETP Lett. {\bf 20} 194 (1974).
\item {4.} E. Amaldi in ``Old and New Problems in Elementary Particles'' ed. G. Puppi
(Acad. Press, New York 1968).
\item {} A.S. Goldhabor and J. Smith, Rep. Prog. Phys. {\bf 38} 731 (1975).
\item {} B. Cabrera, Phys. Rev. Lett. {\bf 48} 1378 (1982)    

\item {} B. Cabrera etal. Phys. Rev. Lett. {\bf 51} 1993 (1983).
\item {} T. Hara et.al. Phys. Rev. Lett. {\bf 56}  553 (1986).
\item {} T. Gentile et.al. Phys. Rev. {\bf D35} 1081 (1987).
\item {} The MACRO collaboration, Nucl. Instrum. Methods. Phys. Res. Sec. A {\bf 264}
 18 (1988).
\item {5.} N. Cabibbo and E. Ferrari, Nuovo Cimento {\bf 23} 1147 (1962).
\item {} C.R. Hagen, Phys. Rev. {\bf 140B} 804 (1965).
\item {} D. Zwanziger, Phys. Rev.  {\bf D3} 880 (1971).
\item {} W. Baker and F. Graziani, Phys. Rev. {\bf D18} 3849 (1978) and
Am. J. Phys. {\bf 46} 1111 (1978).
\item {} D. Singleton, Am. J. Phys. {\bf 64} 452 (1996).
\item {6.} T.W. Hansch, Physics Today {\bf 30}  34 (1977).
\vfill
\eject
\end